\documentclass[sn-mathphys,Numbered]{sn-jnl}

\usepackage{graphicx}%
\usepackage{multirow}%
\usepackage{amsmath,amssymb,amsfonts}%
\usepackage{amsthm}%
\usepackage{mathrsfs}%
\usepackage[title]{appendix}%
\usepackage{xcolor}%
\usepackage{textcomp}%
\usepackage{manyfoot}%
\usepackage{booktabs}%
\usepackage{algorithm}%
\usepackage{algorithmicx}%
\usepackage{algpseudocode}%
\usepackage{listings}%
\usepackage{tcolorbox}
\usepackage{subcaption}
\usepackage[utf8]{inputenc}
\usepackage{stmaryrd}
\usepackage{placeins}

\begin{document}

\title[Article Title]{Time and event symmetry in quantum mechanics}


\author*[1]{\fnm{Michael} \sur{Ridley}}\email{mikeridleyphysics@gmail.com}

\author[2,3]{\fnm{Emily} \sur{Adlam}}\email{eadlam90@gmail.com}

\affil[1]{\orgdiv{School of Physics and Astronomy}, \orgname{Tel Aviv University},  \city{Tel Aviv}, \postcode{69978}, \country{Israel}}

\affil[2]{\orgdiv{Department of Philosophy}, \orgname{Chapman University}, \orgaddress{\street{1 University Drive}, \city{Orange}, \postcode{92866}, \state{California}, \country{USA}}}

\affil[2]{\orgdiv{Department of Physics}, \orgname{Chapman University}, \orgaddress{\street{1 University Drive}, \city{Orange}, \postcode{92866}, \state{California}, \country{USA}}}


\abstract{We investigate two types of temporal symmetry in quantum mechanics. The first type, time symmetry, refers to the inclusion of opposite time orientations on an equivalent physical footing. The second, event symmetry, refers to the inclusion of all time instants in a history sequence on an equivalent physical footing. We find that recent time symmetric interpretations of quantum mechanics fail to respect event symmetry. Building on the recent fixed-point formulation (FPF) of quantum theory, we formulate the notion of an event precisely as a fixed point constraint on the Keldysh time contour. Then, considering a sequence of measurement events in time, we show that both time and event symmetry can be retained in this multiple-time formulation of quantum theory. We then use this model to resolve conceptual paradoxes with time symmetric quantum mechanics within an `all-at-once', atemporal picture. }




\maketitle

\tableofcontents

\section{Introduction}

There is an obvious tension between the geometric notion of time appearing in relativistic spacetime theories and the notion of time used in textbook quantum mechanics (\cite{horwitz_two_1988}). In the former, time enters as one degree of freedom in a spacetime coordinate system necessitated by transformations between reference frames under boosts and rotations (\cite{Einstein1905}), necessitating an eternalist perspective on time. In the latter, time is a measure of the evolution from some initial condition, and enters as a monotonically increasing background parameter in the quantum state (\cite{schrodinger1926undulatory}). This notion of time is closer to that experienced by observers, who have a clear notion of processes involving change, and perceive reality from the presentist perspective.

The two notions of time come with two notions of symmetry. In most discussions, \textbf{time symmetry} refers to a symmetry based on the two different time \emph{orientations} of evolution directed towards the future and the past respectively. However, the geometric picture of spacetime suggests an entirely different notion of symmetry, that of the fundamental physical description accorded to a system defined at different time points, which we term \textbf{event symmetry}. Indeed, the relativity of simultaneity for space-like separated events suggests that a strict ordering of events in time is not present at the level of physical law, and so the separation of times into past, present and future is illusory at that level.    

Time-symmetric formulations of quantum mechanics have been a topic of hot debate since the  retrodictive formulation of  \cite{watanabe_symmetry_1955}. A time-symmetric theory of measurement statistics for pre- and post-selected measurements was then devised by \cite{aharonov_time_1964}. This lead to the two-state vector formalism (TSVF) describing measurements on microscopic systems in terms of both a forwards and backwards-evolving wavefunction defined at the measurement time. In the same year, \cite{keldysh_diagram_1964} presented a formalism for the calculation of ensemble averages in the field theory of quantum statistical mechanics which involved the propagation of states both forwards and backwards in time. Both papers have been extremely influential in their breadth of applicability. The ABL formalism has been applied to the theory of weak measurement (\cite{aharonov_two-state_2008,tamir_introduction_2013}), to the interpretation of measurements in nested interferometers (\cite{vaidman_past_2013}) and to the resolution of paradoxes in quantum theory (\cite{ravon2007three,vaidman2009two}). The Keldysh formalism has been enormously successful as a theory of many-body dynamics, and is the foundational picture for the nonequilibrium Green's function (NEGF) framework that has found many applications in quantum transport, spectroscopy and the thermodynamics of nanoscale systems (\cite{kadanoff_quantum_1962,danielewicz1984quantum,fetter_quantum_2003,van2006introduction,stefanucci_nonequilibrium_2013,cohen_greens_2020,ridley2022many}).

In addition, recent works indicate that \textbf{time symmetry} may actually be a fundamental feature of quantum theory, and either modify the theory to incorporate \textbf{time symmetry} (\cite{sutherland2008causally,price_does_2012}) or argue that the existing theory is fundamentally time symmetric (\cite{di2021arrow}). Moreover, there exist numerous proposals in the literature for retrocausal interpretations of quantum mechanics, which are often motivated by the possibility of rescuing locality from EPR-type arguments \cite{costa1976time,rietdijk1978proof,cramer_transactional_1986,cramer_overview_1988,aharonov1991complete,wharton_time-symmetric_2007,price2008toy,sutherland2008causally,argaman_bells_2010,kastner_transactional_2013,price2015disentangling,sutherland2017retrocausality,cohen2017quantum,adlam2018spooky,friederich2019retrocausality,cohen2020realism}.

However, the principle that we refer to as  \textbf{event symmetry} has not thus far received much attention. In section \ref{eventsymmetry} of this work we discuss the motivations for postulating \textbf{event symmetry} as a natural extension of \textbf{time symmetry}. In sections \ref{TSVF} and \ref{transactional} we discuss two well-known retrocausal approaches to quantum mechanics, the two-state vector approach and the transactional interpretation, and we explain why they do not respect \textbf{event symmetry} as currently formulated. We will see that the absence of \textbf{event symmetry} leads to conceptual difficulties in both cases. Finally, in section \ref{Keldysh} we use these observations to motivate an explicit model which respects both \textbf{time symmetry} and \textbf{event symmetry} based on the recently-proposed Fixed-Point Formulation (FPF) of quantum mechanics (\cite{ridley2023quantum}).

\section{Event symmetry in quantum mechanics \label{eventsymmetry}}

In this article, we will use the term \textbf{event symmetry} to refer to the principle that there is no ontologically privileged moment of time (or in relativistic terms, no ontologically privileged spacelike hyperplane). There are a number of good reasons to think that our scientific theories should respect \textbf{event symmetry}. First of all, one might simply invoke Ockham's razor:  if there is no clear need to privilege some particular moment of spacetime, we should avoid doing so, since the result will be a simpler theory. In addition, the empirical data does not seem to offer compelling evidence for a special role for any moment of time, so if our theories privilege some moments over others, we are introducing an `\emph{asymmetry which does not appear to be inherent in the phenomena}'(\cite{Einstein1905}). It is a time-honoured methodological principle in physics that this should be avoided when possible (e.g. this principle was one of Einstein's main motivations for special relativity). Moreover, \cite{berenstain_privileged-perspective_2020} argues compellingly that we can think of much of the history of science as `\emph{a series of discoveries that have continually
dethroned humankind from a presumed special and unique place in the universe,}' and thus she advocates a form of realism which specifically seeks to avoid postulating privileged perspectives.  And since removing privileged perspectives has been a successful scientific strategy in the past, inductive inference would seem to support applying it also at the level of events: if there are no privileged perspectives, there should not be any privileged events either. 

So the principle of \textbf{event symmetry} seems compelling for a number of reasons. But what does \textbf{event symmetry} actually entail for scientific practice? Minimally, it suggests that the laws of nature should not be time or space-indexed - the same laws should act at every point in space and time. This condition is already met by the vast majority of physical theories that have been proposed: laws which vary with time and/or space are very much in the minority. However, it’s unclear that the principle of \textbf{event symmetry} is upheld at the level of the physical models defined by the laws. In general, it is true within most physical theories that a given experiment is described by the same model regardless of when and where in spacetime that experiment takes place. But within the model, it is not usually true that all moments of time have the same status. For in order to apply the laws, one typically specifies an initial condition (and/or a final condition) and then uses the boundary conditions to calculate states at some other times. Thus structurally the model treats the moments of time at the boundaries as special, so even if the laws don’t violate \textbf{event symmetry}, it appears that the models do.

Now, a natural response to this suggests itself: the boundary conditions should not be regarded as privileged in any deep  sense. They merely reflect our epistemic circumstances - the boundary conditions represent what we currently know, with the states at other times representing what we want to calculate. So the fact that structurally the boundaries appear to play a special role in the models shouldn’t be regarded as implying that the boundaries play a special role at the \emph{ontological} level. 

This argument seems reasonable so long as we only wish to use our models to calculate things. But what if we regard our models as attempts to represent the way things really are? Then, if we really believe in \textbf{event symmetry}, we should surely believe that it is possible to write down models in which the equal status of all moments of time is transparent: that is to say, models which are structured in such a way that no moment of time is treated differently from any other moment of time. The kinds of models we actually use for calculations, which have distinct `inputs/boundary conditions’ and `outputs/predictions’ can then be regarded as derived from the underlying models representing the real structure of the world: these input-output models are a useful tool when one wants to use the models for prediction, but the input-output structure is purely epistemic and does not directly represent the structure of the world, which is to be found in the underlying model that does not privilege any moment of time. In this article we will refer to models intended purely for calculation as `calculational models' - such models should not be expected to obey \textbf{event symmetry} - and we will refer to models intended as representations of reality as `representational models,' with the expectation that representational models should ultimately obey \textbf{event symmetry}. 

Are there any examples of representational models which indeed obey \textbf{event symmetry}? As a matter of fact, any deterministic and reversible time-evolution model can be written in this way. As an example, take Newtonian mechanics or unitary quantum mechanics. Usually we apply such models by selecting an initial condition and evolving forwards in time. But we could equally well select any other time-slice and evolve both forwards and backwards from it. So if we insist on a  picture which starts at some particular moment of time and treats that choice as having representational content, we will be faced with massive underdetermination, since we can never determine empirically which time-slice is really the privileged time-slice. However, this can easily be avoided if we adopt a picture which does not start from any moment of time.  As detailed in ref \cite{adlam2022laws}, one could simply specify a set of `dynamically possible histories,’ and then rather than choosing an initial condition one could simply sample from the set of possible histories. This kind of `all-at-once' model, which simply assigns probabilities over entire histories, transparently obeys \textbf{event symmetry}. 

From this point of view, the usual calculational models of Newtonian mechanics or unitary quantum mechanics, which take initial conditions as inputs, can be regarded as being derived from underlying representational models which specify `dynamically possible histories’ - thus the special role of the initial conditions has only epistemic significance, and doesn’t directly represent anything in the structure of the world. In this sense, the `arbitrariness’ associated with selecting an initial condition in a standard time-evolution model need not really be regarded as being located in any specific spacetime location, but rather as delocalised over the whole course of history. In  \cite{adlam2022determinism} this condition is formalised under the name of `delocalised holistic determinism.’

It should be noted that this feature of deterministic, reversible models is not always reflected in the way we reason about these theories. When we think about things in cosmological terms, for example, the initial condition of the universe is almost invariably treated differently - all moments of time are equal, but some moments are more equal than than others!  In particular, the second law of thermodynamics is typically explained by making a specific postulation about the initial state of the universe (see \cite{Price2004-PRIOTO-2,pittphilsci8894}). But the mathematical form of these theories offers no grounds whatsoever for the idea that everything must be explained by appeal to the initial state. The justification for this approach seems to come from an \emph{interpretation} of the theories which regard the universe as something like a computer which takes an initial state and evolves it forwards in time (see \cite{wharton2015universe}) - but since nothing in the theory singles out the initial state in this way, this interpretation seems to be based on prior assumptions rather than justified by the theory itself. Indeed, the main reason why we tend to describe systems as starting at some initial state and evolving forward in time is presumably because we ourselves experience time as progressing forwards and we are inclined to project our experience onto reality. But it has long been argued that our experience of temporal progression may not reflect any objective process of temporal becoming (\cite{Mellor1981-MELRT-2,Dyke2007-DYKMAT-2}), so we should beware of fallacies that arise from inappropriately generalising the temporal nature of our experience.

Now, it is important to note that re-interpreting a deterministic and reversible time-evolution model is not the only possible way of arriving at a representational model obeying \textbf{event symmetry} -  as  noted by  \cite{adlam2022laws}, once we move from a time-evolution picture to an all-at-once picture there are a variety of new possibilities open to us, since we can also contemplate possible all-at-once models which \emph{cannot} be written in a time-evolution form.   In particular, one might expect to find representational models which exhibit some kind of nontrivial dependence of the future on the past, something which would look like retrocausality if one attempts to understand it within a time-evolution picture. So it seems natural to think that various retrocausal formulations of quantum mechanics might perhaps yield further representational models obeying \textbf{event symmetry}. 

However, much will depend on the kind of retrocausality instantiated by the relevant models. \cite{adlam2022roads} observes that the literature includes two quite different conceptions of retrocausality.   `Dynamical retrocausality' postulates two distinct directions of dynamical causality which together determine intermediate events by forwards and backwards evolution respectively from separate and independent initial and final states - for example, the forwards-evolving state and the backwards-evolving state in the two-state vector interpretation \citep{aharonov_time-symmetric_2010-1}. Evidently models of this kind, if regarded as representational models, do not respect \textbf{event symmetry} since they privilege the initial and final states. Alternatively, `All-at-once' retrocausality suggests that the laws of nature apply atemporally to the whole of history, as for example in Wharton's all-at-once Lagrangian models \citep{Wharton_2018}; in such a picture the past and the future have a reciprocal effect on one another, so there is definitely some kind of influence from the future to the past at play, but these effects can't be separated out into distinct forwards and backwards evolutions. In general we would expect models of this kind to obey \textbf{event symmetry}, since there is no special point in time from which the evolution begins. So if we are seeking models which obey \textbf{event symmetry}, we are clearly better off working with all-at-once models.

 Indeed,  as argued  by \cite{adlam2022roads}, dynamical retrocausality is very difficult to implement in a consistent way - it tends to lead to contradictions like the grandfather paradox, which can only be resolved by applying consistency conditions that then have to be understood within an all-at-once description in any case. It also leads to objections of the form of the criticism of the transactional interpretation given by \cite{maudlin2011quantum}, which we discuss in more detail in section \ref{transactional}. On the other hand, all-at-once retrocausality is quite straightforward to implement, by means of laws of nature which specify constraints on what is physically possible. Laws of this form have been taken seriously in physics for some time (\cite{Deutsch_2015,Filomeno2021-FILTOD}) and have recently  attracted  philosophical attention (\cite{adlam2022laws,chen2022governing}). So even aside from \textbf{event symmetry}, there are good reasons to prefer all-at-once retrocausal models over dynamical retrocausal models. 

Now, it should be noted that there is a sense in which any `all-at-once’ theory can be regarded as retrocausal: since we are no longer postulating evolution which `begins’ at the start of time, or indeed at any particular moment of time, it follows that every moment depends mutually and reciprocally on any other moment, and therefore past moments depend on future moments in just the same way that future moments depend on past moments. However, the kind of retrocausality involved in simply re-interpreting Newtonian mechanics or unitary quantum mechanics in an all-at-once way is not particularly interesting: ideally we would like to find models exhibiting a more novel kind of retrocausality, i.e. retrocuasality which is \emph{irreducible} in the sense that  the dependence of the past on the future can't be removed by simply rewriting the model in terms of initial conditions and a time-evolution law. A model of this kind would postulate a distribution of probabilities over entire courses of history which has the property that it cannot be rewritten as a distribution of probabilities over initial states plus an evolution law. For example, there exist several proposals for `retrocausal' interpretations of quantum mechanics which exhibit this kind of irreducible retrocausality; we will discuss two of these approaches now, before presenting  a new retrocausal model based on the Keldysh contour in section \ref{Keldysh}.

\section{Transactional Interpretation \label{transactional}}

The Transactional Interpretation (TI) of Cramer \cite{cramer_transactional_1986,cramer_overview_1988, kastner2016transactional} is a version of quantum theory inspired by the time-symmetric formulation of electromagnetism by Wheeler and Feynman \cite{wheeler1945interaction}. In the Wheeler-Feynman theory, both advanced and retarded solutions of Maxwell's equations are incorporated equivalently. The electromagnetic field emitted by a charged
particle is set to half the
sum of the retarded and advanced solutions of Maxwell’s
equations. The emitted wave then interacts with other particles, the so-called `absorber', which respond with half advanced, half retarded waves, and the combination of the emitter and absorber fields leads to processes in which energy is transmitted from emitter to absorber. Similarly, Cramer puts both the standard (forwards time) solutions to Schrodinger equation and their complex conjugates (backwards time) on an equivalent ontological footing. The usual quantum state, (denoted as a `ket' vector), propagating from the emitter is called the `offer' wave. This is combined with the advanced response (denoted with a `bra' vector) from the absorber, called the `confirmation' wave. The two-way process between emitters and absorbers is called a `transaction', hence the TI.

\begin{figure}
\centering
  \includegraphics[width=.5\linewidth]{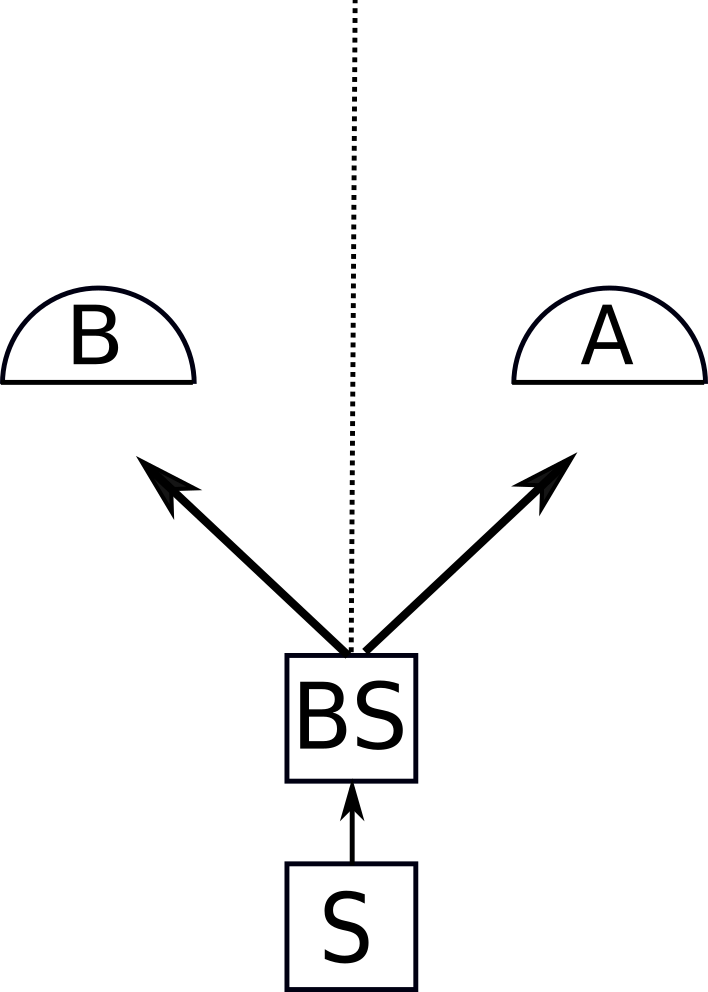}
  \caption{The basic interferometric setup used in the discussion of the TI.}
  \label{fig:Maudlin_1}
\end{figure}

To illustrate the TI, we apply it to the experiment shown in Fig. \ref{fig:Maudlin_1}. A particle is emitted by a source S at time $t_{0}$, passes through a symmetric beam-splitter BS, and is sent to either the right, where it is absorbed by detector A, or to the left, where it is absorbed by B. The particle detection event occurs at time $t_{1}$ in either case. The TI states that the source emits a retarded offer wave at $t_{0}$ in the direction of BS, where it is separated into two equal-amplitude components. The arrival of an offer wave at A or B at $t_{1}$ triggers the emission of an advanced confirmation wave of the same amplitude and phase travelling back in time to the source at $t_{0}$. Therefore, there exists two possible equal-amplitude offer-confirmation pairs between the emitter and the absorbers, one going to A and one to B. If the completed transaction occurs between the source and A, then the particle is detected at A, and vice versa for B. The chance of each possible transaction occurring is given by the product of the offer wave and confirmation wave amplitudes, leading to a probability of 1/2 for each completed transaction. 

Evidently the TTI in the formulation we have just given fails to obey \textbf{event symmetry}: the time of the emission and the time of the absorption play the role of boundary conditions, and hence these moments of time, and that of the measurement, are specially privileged in the construction. However,  there is potentially a way out of this dilemma, because one could imagine a version of the model which does not treat these moments as special but rather postulates a general law applying to all moments of time such that, if some particular condition is met at that moment, an offer wave is emitted forwards in time, and likewise another law applying to all moments of time such that, if some particular condition is met at that moment, a confirmation wave is emitted backwards in time. This model would obey \textbf{event symmetry} because the emission and absorption would be the result of physical laws applying equally to all moments of time, rather than ad hoc boundary conditions. However, in order to make this work it would be necessary to come up with a clear physical condition defining events leading to emission (i.e. state preparation) and absorption (i.e. measurement) and this is non-trivial - the fact that we cannot easily give a clear physical condition defining `measurements' is in many ways the essence of the measurement problem. So although it is possible to imagine an event-symmetric formulation of the TI if emission and absorption can be regarded as contingent physical events rather than artificially imposed boundary conditions, there are complexities involved in actually implementing such a thing.

\subsection{Maudlin's challenge}

\begin{figure}[!ht]
  \begin{subfigure}{\linewidth}
  \centering
    \includegraphics[width=.35\linewidth]{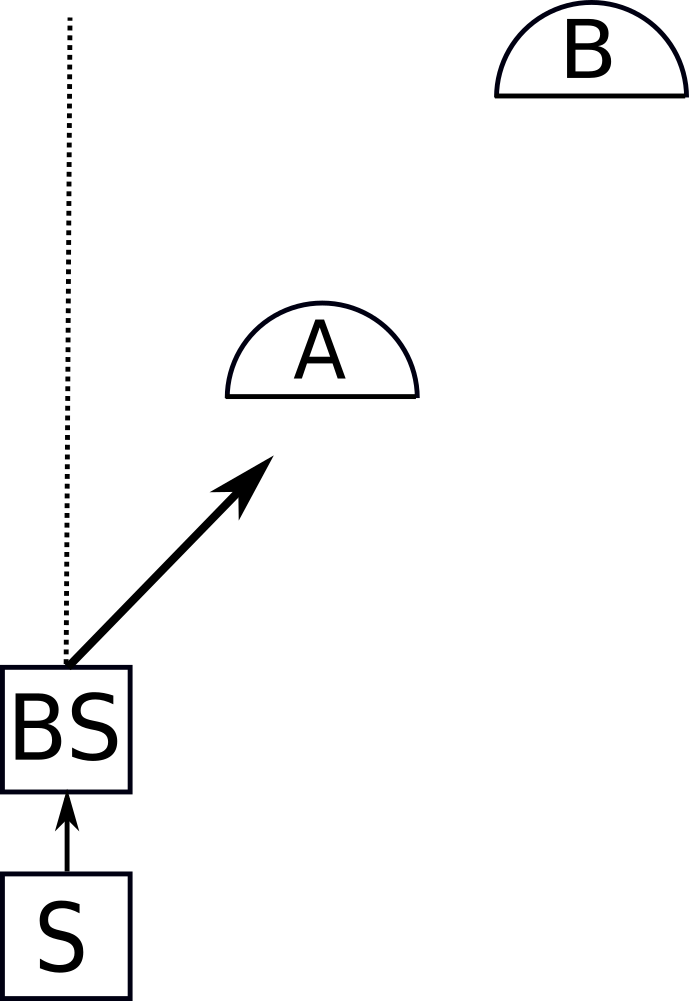}
    \subcaption{}
  \end{subfigure}
  
  \begin{subfigure}{\linewidth}
    \centering
    \includegraphics[width=.5\linewidth]{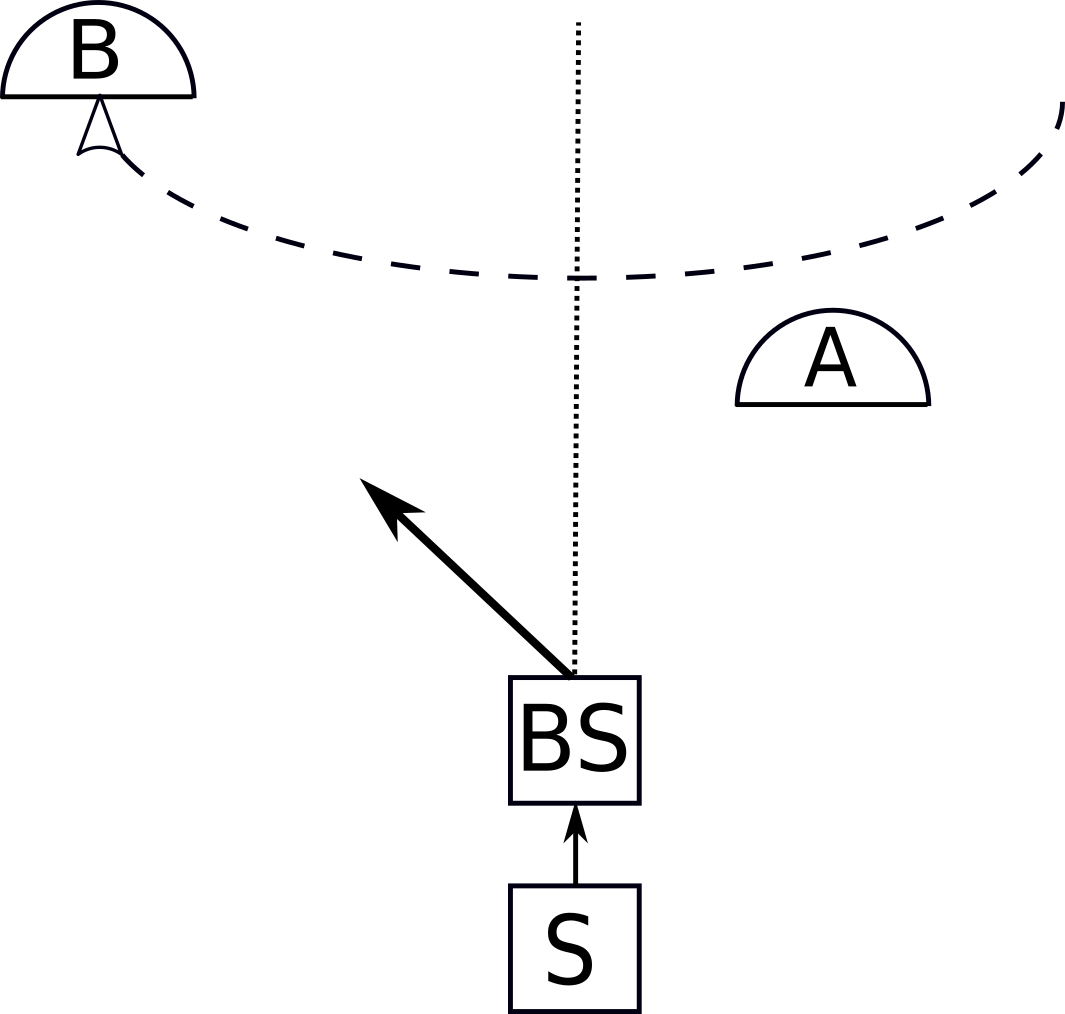}
    \subcaption{}
  \end{subfigure}
  \caption{Schematic of the setup in Maudlin's challenge, in the cases (a) a photon is emitted to the right-hand side of the beam splitter to be detected at A before later being detected at B and (b) a photon is emitted to the left-hand side of the beam splitter such that there is no detection event at detector A, which causes the detector B to swing around to the left-hand side and intercept the incoming particle.}\label{Maudlin_plots}
\end{figure}

Maudlin has proposed an objection to retrocausal quantum theories, and in particular to the TI \cite{maudlin2011quantum}, which makes use of causal loops. Consider the experiment shown schematically in Fig. \ref{Maudlin_plots}. A particle is emitted at S at time $t_{0}$, and separated into two equal-amplitude waves by the beam-splitter BS. If it goes to the right (Fig. \ref{Maudlin_plots} (a)) it is detected by absorber A at time $t_{1}$. If it goes to the left (Fig. \ref{Maudlin_plots} (b)) then A does not detect a particle at $t_{1}$, which triggers a detector at B to be swung from its initial position behind A to a point on the left of BS, where it detects the particle at a later time $t_{2}$. If BS produces a particle with equal amplitudes in the left and right directions, then according to standard quantum mechanics there is a probability 1/2 to detect the particle at detector A on the right side, and at B on the left side of the beam splitter, respectively.

Maudlin's challenge may be stated thus: the equiprobability of the two outcomes demands that equal amplitude confirmation waves be received from absorber A and absorber B. But if B sends back a confirmation wave, then it must have been struck by an offer wave, so it must have swung round. And if it has swung round, then the particle is not detected at A. So when a confirmation wave is received from absorber B, the particle must go to absorber B with probability 1, despite the fact that the offer-confirmation amplitude product is 1/2.

The inconsistency arises because the configuration of the absorbers is contingent—the location of B depends on what happens at A—whereas the TI takes the configuration of absorbers as fixed boundary conditions. If the absorbers aren't fixed, the recipe yields inconsistent results.
Maudlin's challenge is an effective criticism against the transactional interpretation of quantum mechanics because of the latter's emphasis upon the dynamical process of sending out the offer and confirmation waves. Maudlin's challenge is a problem for this type of retrocausality, but not necessarily for the all-at-once approach. And of course, as we have already noted, an all-at-once formulation is also much more likely to be compatible with \textbf{event symmetry}. Thus in Section \ref{Keldysh} we construct an all-at-once model of retrocausality  and show how it resolves Maudlin's challenge.

\section{Two State Vector Formalism \label{TSVF}}

One well-known approach to retrocausality in quantum mechanics is the two-state vector formalism, in which one defines both an initial state and a final state and the probabilities for intermediate events are determined by both the initial and final state. Clearly this kind of model exhibits a form of \textbf{time symmetry}, but does it also exhibit \textbf{event symmetry}? In fact there are reasons to think that it does not, for as ref \cite{adlam2022roads} observes, the two-state vector formalism is most naturally interpreted as a form of dynamical retrocausality, rather than all-at-once retrocausality, and we have already noted that dynamical retrocausality does not seem very compatible with \textbf{event symmetry}. 

Moreover, not only is the two-state vector formalism naturally interpreted in a dynamical way, it is unclear that there is any possible way to translate it into an all-at-once model. We can’t simply rewrite the two-state vector formalism as a rule specifying sets of `dynamically possible histories’ - for if we just apply the standard quantum-mechanical evolution law to say which `forwards’ and `backwards’ histories are possible, then we will end up with the same histories in both directions, i.e. the histories which are allowed by standard quantum mechanics. So attempting to rewrite the two-state vector interpretation as a representational model in a form which does not privilege any moment of time just seems to give us back standard quantum mechanics. Thus it seems that the two-state vector interpretation relies crucially on treating certain moments of time differently, which means it most likely cannot respect the principle of \textbf{event symmetry}. We will now see how this plays out in the context of several different versions of the two-state formalism. 

\subsection{The two state vector for a pair of times}\label{TSVF_pair}

The basic experimental set-up considered in the TSVF is that of a system defined between two fixed points in time, $t_{1}$ and $t_{2}>t_{1}$, at which strong measurements are carried out on the system. This results in

1. the pre-selected state $\left|\psi\left(t_{1}\right)\right\rangle$  which travels forwards in time through the time interval $\left[t_{1},t_{2}\right]$ in accordance with the TDSE $i\partial_{t}\left|\psi\left(t\right)\right\rangle =H\left(t\right)\left|\psi\left(t\right)\right\rangle$  and

2. the post-selected state $\left\langle \phi\left(t_{2}\right)\right|$ which travels in the backwards time direction in accordance with the conjugate TDSE $-i\partial_{t}\left\langle \phi\left(t\right)\right|=H\left(t\right)\left\langle \phi\left(t\right)\right|$.

Now suppose a measurement is carried out at the intermediate time $t\in\left[t_{1},t_{2}\right]$. The preselected state $\left[\psi\left(t_{1}\right)\right\rangle$  evolves from $t_{1}$ to the measurement time $t$ in accordance with the unitary evolution operator
\begin{align}
    U\left(t,t_{1}\right)=\hat{T}\left\{ e^{-i\int_{t_{1}}^{t}H\left(\tau\right)d\tau}\right\},
\end{align}
where $\hat{T}$ is the chronological time-ordering operator.

The postselected state is evolved back in time from $t_{2}$ $t$ in accordance with the evolution
\begin{align}
    U\left(t_{2},t\right)=\tilde{T}\left\{ e^{-i\int_{t}^{t_{2}}H\left(\tau\right)d\tau}\right\}, 
\end{align}

where $\tilde{T}$ is the anti-chronological time-ordering operator. The two oppositely-orientated parts of the system can then be combined into a single `two state vector'

\begin{equation}\label{eq:TSV2}
    \left\langle \phi\left(t_{2}\right)\right|\otimes\left|\psi\left(t_{1}\right)\right\rangle, 
\end{equation}

which exists in the composite Hilbert space constructed from distinct time-localized `universes' existing at single times \cite{aharonov_each_2014}

\begin{equation}
    \mathcal{H}_{t_{2}}^{\dagger}\otimes\mathcal{H}_{t_{1}}
\end{equation}

States in this Hilbert space are fundamentally (i) time non-local objects and (ii) built out of parts with opposite time orientations. The solution to understanding the apparent time asymmetry in quantum mechanics, according to the TSVF, is thus to revise the concept of a quantum state to include two time degrees of freedom (\cite{reznik1995time}). We note, however, that the two-state vector is not a state vector composed out of multiple degrees of freedom in the usual sense, but is rather an operator constructed from a vector and its dual (located in the dual Hilbert space).

\begin{figure}[!ht]
  \begin{subfigure}{\linewidth}
  \centering
    \includegraphics[clip, width=.75\linewidth]{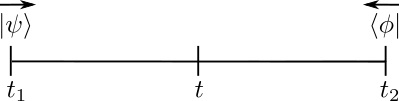}
    \subcaption{}
  \end{subfigure}%
  
  \begin{subfigure}{\linewidth}
  \centering
    \includegraphics[clip, width=.75\linewidth]{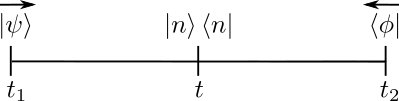}
    \subcaption{}
  \end{subfigure}%
\caption{TSVF picture of complete measurements at $t_{1}$ and $t_{2}$ with (a) only weak measurements in between and (b) with a third complete measurement at the intermediate point $t\in\left[t_{1},t_{2}\right]$. Arrows indicate direction of time-propagation.} \label{fig:TSVF}
\end{figure}

Fig. \ref{fig:TSVF} (a) shows the time segment $\left[t_{1},t_{2}\right]$ with a weak measurement performed at t. This measurement does not affect the state at that time but instead couples it very weakly to a variable $\hat{O}$, with the corresponding weak value:
\begin{align}
    O_{w}\equiv\frac{\left\langle \phi\left(t_{2}\right)\right|U\left(t_{2},t\right)\hat{O}U\left(t,t_{1}\right)\left|\psi\left(t_{1}\right)\right\rangle }{\left\langle \phi\left(t_{2}\right)\right|U\left(t_{2},t\right)U\left(t,t_{1}\right)\left|\psi\left(t_{1}\right)\right\rangle }.
\end{align}

In Fig. \ref{fig:TSVF} (b), we show the same time segment as in (a) but this time with an additional strong measurement of the state $\left|n\right\rangle \in \left\{ \left|k\right\rangle \right\} $  at the time $t$. According to the TSVF, to obtain the probability of measuring the system in some state $\left|n\right\rangle \in \left\{ \left|k\right\rangle \right\}$ at the intermediate time $t \in \left[t_{1},t_{2}\right]$, the system is propagated in \emph{both} time directions, from $t_{1}\rightarrow t$ and $t_{2}\rightarrow t$, such that the amplitude of the $n$-th outcome is given by sandwiching this state between the forwards and backwards-oriented parts of Eq. \eqref{eq:TSV2}:

\begin{equation}
    \left\langle \phi\left(t_{2}\right)\right|U\left(t_{2},t\right)\left|n\right\rangle \left\langle n\right|U\left(t,t_{1}\right)\left|\psi\left(t_{1}\right)\right\rangle 
\end{equation}

Then, assuming the Born rule, the normalized modulus-square of this yields the probability to obtain outcome $n$:

\begin{equation}\label{ABL}
    P_{n}=\frac{\left|\left\langle \phi\left(t_{2}\right)\right|U\left(t_{2},t\right)\left|n\right\rangle \left\langle n\right|U\left(t,t_{1}\right)\left|\psi\left(t_{1}\right)\right\rangle \right|^{2}}{\underset{k}{\sum}\left|\left\langle \phi\left(t_{2}\right)\right|U\left(t_{2},t\right)\left|k\right\rangle \left\langle k\right|U\left(t,t_{1}\right)\left|\psi\left(t_{1}\right)\right\rangle \right|^{2}}
\end{equation} 

This is known as the ABL rule. It was obtained by taking quantum states built up out of moments of time, each of which corresponds to an independent state space. 

The above account of multiple time measurements makes accurate predictions, but if one considers it as a representational model one may ask why the state of the system only has a forward-directed component at $t_{1}$ and a backward-directed state at $t_{2}$? What is the nature of the backward-directed part? If it is ontic, where is the backwards-directed part of the state at $t_{1}$ and the forwards-directed part at $t_{2}$? 

To illustrate the problem this causes, we may consider the intermediate time $t \in \left[t_{1},t_{2}\right]$ and ask: without any measurement, what is the state of the system at $t$? Propagating from $t_{1} \longrightarrow t$ gives $U\left(t,t_{1}\right)\left|\psi\left(t_{1}\right)\right\rangle$, whereas propagating from $t_{2}\longrightarrow t$ will give $\left\langle \phi\left(t_{2}\right)\right|U\left(t_{2},t\right)$ for the state of the system. The TSVF appears to give two answers, one of which exists in $\mathcal{H}_{t}$, the other in $\mathcal{H}_{t}^{\dagger}$. This opens the possibility that the state at any time $t \in \left[t_{1},t_{2}\right]$ has \emph{two} components, oriented in opposite directions of time, such that the initial two-state vector is mapped to an object in the composite Hilbert space $\mathcal{H}_{t}^{\dagger}\otimes\mathcal{H}_{t}$. The two parts of this state are defined in distinct Hilbert spaces and, strictly speaking, cannot overlap.

Furthermore, if we allow the states to \emph{cross} at time $t$, there is an apparent ambiguity about the definition of the two-vector at any pair of times $t',t" \in \left[t_{1},t_{2}\right]$, with $t">t'$, $t' \in \left[t_{1},t\right]$ and $t" \in \left[t,t_{2}\right]$. It can either be given by  
\begin{equation}\label{eq:TSV}
    \left\langle \phi\left(t_{2}\right)\right|U\left(t_{2},t"\right) \otimes U\left(t',t_{1}\right)\left|\psi\left(t_{1}\right)\right\rangle, 
\end{equation}
or
\begin{equation}
    \left\langle \phi\left(t_{2}\right)\right|U\left(t_{2},t'\right) \otimes U\left(t",t_{1}\right)\left|\psi\left(t_{1}\right)\right\rangle. 
\end{equation}

\begin{figure}
\centering
    \includegraphics[clip, width=.75\linewidth]{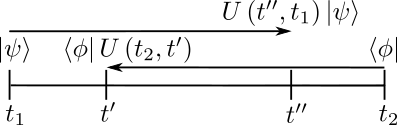}
    \caption{ Arrows indicate direction of time-propagation.}
    \label{fig:crossing}
\end{figure}

The latter case defines the case of \emph{crossing}, shown in Fig. \ref{fig:crossing}. This process is only possible because there is no additional boundary condition introduced for the intermediate time $t\in\left[t',t''\right]$. Thus there appears to be a lack of clarity about the specification of the two-state vector at the pair of times $t',t"$.

We may also consider the case in which crossing is prevented, but a strong measurement at time $t$ puts the system into the state $\left|n\right\rangle \in\left\{ \left|k\right\rangle \right\} $. We assume that both time components of this state are given by $\left|n\right\rangle$, i.e. the two-state vector here is $\left\langle n \right| \otimes \left|n\right\rangle \in \mathcal{H}_{t}^{\dagger}\otimes\mathcal{H}_{t}$. then we ask: what is the two state vector at the pair of times $t', t"$?

To clarify the concepts one may invoke to answer this question, we introduce the notion of a \emph{fixed point state} at time $t$:

\textbf{Fixed Point at time $t$} - \emph{A state with equal forwards and backwards components defined at the time $t$.}

\begin{figure}
  \begin{subfigure}{\linewidth}
  \centering
    \includegraphics[clip, width=.75\linewidth]{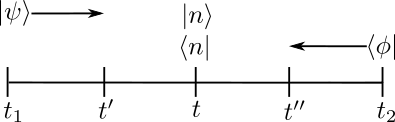}
    \subcaption{BTFP propagation.}
  \end{subfigure}%
  
  \begin{subfigure}{\linewidth}
  \centering
    \includegraphics[clip, width=.75\linewidth]{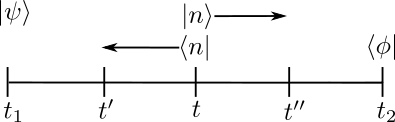}
    \subcaption{FPTB propagation.}
  \end{subfigure}%
\caption{The two types of propagation resulting in a two-time state at the pair of times $t',t"$ are shown, with (a) Boundary-to-fixed-point (BTFP) and (b) Fixed-point-to-boundary (FPTB) propagations illustrated schematically. Arrows indicate direction of time-propagation.} \label{fig:FP}
\end{figure}

There are thus two routes for getting to the pair of times $t',t"$:

(a) Boundary-to-fixed-point (BTFP). Shown in Fig. \ref{fig:FP} (a), this results in the two-state vector 

\begin{equation}
    \left\langle \phi\left(t_{2}\right)\right|U\left(t_{2},t"\right) \otimes U\left(t',t_{1}\right)\left|\psi\left(t_{1}\right)\right\rangle, 
\end{equation}

(b) Fixed-point-to-boundary (FPTB). Shown in Fig. \ref{fig:FP} (b), this results in the two-state vector 

\begin{equation}
    \left\langle n \right|U\left(t,t'\right) \otimes U\left(t",t\right)\left|n\right\rangle, 
\end{equation}

Note that the time ordering of the forwards and backwards time components switches between these two types of propagation. Whereas the TSVF focuses on BTFP propagation, consistently applying \textbf{event symmetry} to each moment of time we would expect \emph{both} BTFP and FPTB propagation to be present.  

So we have shown how, if taken as a representational model of reality, the two-state vector formalism results in paradoxes, or at least conceptual difficulties which can be attributed to the lack of \textbf{event symmetry} in the formalism. These difficulties might be resolved if we instead build the two-time wavefunction as a more complex object with two temporal degrees of freedom at \emph{each} time, one for each time orientation. Before we do this explicitly, we will now consider sequences of events of length $N_{t}>2$.


\subsection{The multiple-time state formalism}\label{ETNU}

The TSVF was generalized to a multiple-time state version, in which any number of vectors propagating from the past and future can be considered in generalized multiple-time states (\cite{aharonov_multiple-time_2009}). Recently, this has been generalized to a novel theory of time propagation itself, the so called `each instant of time a new universe' (ETNU) framework by \cite{aharonov_each_2014}. This formalism describes the wavefunction across regions of time, constructed from wavefunctions defined on tiny time `bricks'. 

\begin{figure}[htp]
\centering
    \includegraphics[clip, width=.75\linewidth]{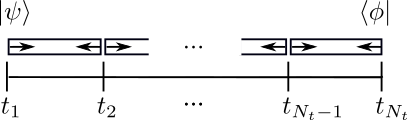}
    \caption{The multiple-time generalization of the TSVF within the ETNU philosophy. Arrows indicate direction of time-propagation.} 
    \label{fig:ETNU}
\end{figure}

Within ETNU, two Hilbert spaces are defined at each time, one for each time orientation. If time is discretized into $N_{t}$ points bounding $N_{t}-1$ time bricks separating a pre-selection of $\left|\psi\right\rangle $ at $t_{1}$ and a post-selection of $\left\langle \phi\right|$ at $t_{N_{t}}$, we thus have $2N_{t}$ Hilbert subspaces, one for each boundary of each brick. The later and earlier brick boundaries correspond to backwards and forwards-directed states, as shown in Fig. \ref{fig:ETNU}. Multiple-time wavefunctions describing the full process are then defined on the following composite Hilbert space:

\begin{equation}\label{eq:ETNU}
    \mathcal{H}\left(N_{t}\right)\equiv\mathcal{H}_{N_{t}}\otimes\mathcal{H}_{N_{t}}^{\dagger}\otimes...\otimes\mathcal{H}_{t_{1}}\otimes\mathcal{H}_{t_{1}}^{\dagger}.
\end{equation}

Time evolution is then replaced with correlations between subsequent time moments in the multiple-time state

\begin{equation}
    \left\langle \phi\right|\underset{i}{\sum}U_{t_{N_{t}}t_{N_{t}-1}}\left|i\right\rangle _{t_{N_{t}}}\left\langle i\right|_{t_{N_{t}-1}}\ldots\underset{i}{\sum}U_{t_{2}t_{1}}\left|i\right\rangle _{t_{2}}\left\langle i\right|_{t_{1}}\left|\psi\right\rangle,
\end{equation}

where $U_{t_{2}t_{1}}$ is numerically equal to the unitary evolution operator $U\left(t_{2},t_{1}\right)$. \cite{aharonov_each_2014} then `contract' the vectors belonging to the past and future boundary conditions, i.e. they allow the bras and kets in the above expression to simply overlap, generating the amplitudes of large temporal sequences of events.

Because the multiple-time formalism allows us to have vectors propagating forwards and backwards from many different origin points, it appears to do better than the two-time picture with respect to implementing \textbf{event symmetry}.  However, there are a number of technical problems with this formalism. First, it doesn't really make sense that the forwards and backwards travelling parts of the two-state vector are  permitted to overlap - they exist in distinct Hilbert spaces and should be kept distinct. In particular, it should not be allowed to `contract' the multiple-time state vector in the ETNU theory to get probability amplitudes, as the forward/backward vectors at the boundaries of the time `bricks' in that approach lie in distinct Hilbert spaces. The two time directions are distinct degrees of freedom and have to be treated as such. For the same reason, we also cannot simply insert unitary propagators to connect bras and kets, in the second type of ETNU contraction discussed above. In section \ref{Keldysh} we will suggest a way to solve these problems, thus arriving at a model which does successfully implement \textbf{event symmetry}. 

We have thus far shown that (i) the TSVF and its multiple-time generalization has mathematical problems and (ii) taken as a representational model, this formalism raises severe conceptual questions. We now move on to a recent interpretational framework for the TSVF and the multiple-time state formalism, and evaluate it in the context of our discussion of \textbf{event symmetry} in Section \ref{eventsymmetry}.

\subsection{The Two Time Interpretation, and why it fails}

Of course, it is possible to consider the formalisms we have just described as merely calculational rather than representational models, in which case their failure to obey \textbf{event symmetry} need not worry us. But in recent years the proponents of this formalism have explicitly supplemented it with an interpretation which indicates that they do wish to regard it as a representational model. This `two-time interpretation,' (TTI), is intended as a solution to the measurement problem, and its adherents claim that it is local, deterministic and ontological (\cite{aharonov2005two,aharonov_measurement_2014,aharonov_two-time_2017,cohen2017quantum}).

According to the TTI, there really exists a backward-travelling wavefunction from the future from a fixed final condition (\cite{cohen2017quantum}).  Thus in the TTI, the two-time state is a direct representation of reality, so the model has been described as `two-psi-ontic' \cite{aharonov_two-time_2017}. On this account, the universe is fundamentally a pre- and post-selected measurement. In addition to the `history vector' $\left|\Psi_{HIS}\right\rangle$ which propagates from the beginning of time, there is a `destiny vector' $\left\langle \Phi_{DES}\right|$ that evolves to the present from the future, and which completely determines the outcomes of measurements in the present so that their statistics follow the Born measure. 

In other words, the final state of the universe is selected to give the right measurement statistics, which begs the question as to why these statistics and no other (see discussion in \cite{robertson2017can}). The TTI thus relies on our experience of the actually-measured statistics to make a convenient selection of the final condition $\left\langle \Phi_{DES}\right|$, similarly to the manner in which Bohmian approaches assume the Born measure in their initial conditions \cite{valentini2020foundations}.

As discussed in Section \ref{eventsymmetry}, the existence of privileged boundary conditions, at which the state of the universe is specially selected to give a certain answer, is clearly incompatible with \textbf{event symmetry}.  Thus this version of the TTI is acceptable if the aim is only to provide a calculational model, but is problematic if regarded as a representational model, insofar as one agrees that representational models should obey \textbf{time symmetry}.

The analysis in Section \ref{TSVF_pair} shows that the existence of fixed point boundary conditions at the measurement time raises problems for the TSVF, although they seem to be implied by that framework, This is perhaps why their existence is explicitly denied by \cite{aharonov_measurement_2014}, where it is stated:

\emph{"A measurement, as empirically observed, generally yields a new outcome state of the quantum system and the measuring device. This state may be treated as an effective boundary condition for both future, and as indicated by the ABL rule, past events. We suggest it is not the case that a new boundary condition is independently generated at each measurement event by some unclear mechanism."}

This claim certainly appears to be making ontological assertions about reality, so it should presumably be interpreted as pertaining to representational models rather than merely calculation ones. Viewed in that way, it clearly violates \textbf{event symmetry}, as it postulates models which use different mathematical representations for different times. In a representational model which obeys \textbf{event symmetry}, the fixed point structure of measurement events must be incorporated into the theory. Moreover in each time `brick' of the multiple-time state, \emph{both} BTFP \emph{and} FPTB propagation must occur. There is no reason to separate the universe into forwards and backwards travelling parts around the `present', resulting from unknown `history' and `destiny' boundary conditions which effectively subject the whole of time to a BTFP-style propagation.

In the next section we will use these ideas to motivate our construction of  a truly time symmetric and event symmetric representational model of reality.

\section{Quantum States in Keldysh Hilbert Space \label{Keldysh}}

We begin by restating the goal - we wish to arrive at a representational model for quantum mechanics which respects \textbf{event symmetry}. In addition, in this article we will aim to do this with a model with no ontology other than the wavefunction (though we do not rule out the possibility that other types of models could also respect \textbf{event symmetry}). Given the previous discussion, we expect that this model would lend itself to an atemporal, `all-at-once' picture exhibiting all-at-once retrocausality.

Our contention is that conceptual issues arise with the TSVF and its multiple-time generalization because it does not consistently apply \textbf{event symmetry}. So here we aim to develop a version of this model which does exhibit \textbf{event symmetry}. To achieve this, we suggest that in each time `brick' of the multiple-time state, \emph{both} BTFP \emph{and} FPTB propagation must occur. Similarly, there is no reason to separate the universe into forwards and backwards travelling parts around the `present', resulting from unknown `history' and `destiny' boundary conditions. This effectively subjects the whole of time to a BTFP-style propagation.

The consistent application of \textbf{event symmetry} to the TSVF does however, give us an important lesson in the construction of a representational time-symmetric model for quantum mechanics. At each moment of time, the quantum state possesses two time directions. These two directions run in parallel, i.e. there should be two time branches with a separate dynamics governing each branch. Each backwards-oriented vector component is connected only to other backwards-oriented vectors by the unitary time evolution, and forwards-oriented vectors are connected to other forwards-oriented vectors. By allowing $U_{t_{2}t_{1}}$ to couple forwards and backwards-oriented vectors, Aharonov \emph{et al.} threw away half the information which one would expect to see in a multiple-time state connecting two time branches. This is a perfectly reasonable thing to do if the aim is to produce a calculational model, in which case it makes sense to throw away information which is not relevant for predicting the target of the model,  i.e. a measurement outcome occurring in between time $t_1$ and $t_2$. But if the aim is to construct a representational model then clearly this information should not be thrown away: it is important to understand how both time directions are integrated at every point in time, not merely at times in between $t_1$ and $t_2$.

Is there a straightforward way to construct a representional model with two time directions included symmetrically? Fortunately in 1964, Keldysh formulated a convenient time-dependent theory for dealing with nonequilibrium statistical dynamical processes. The fundamental idea of the Keldysh method is the use of a time contour with two branches, one oriented forwards in time, one backwards. This idea was recently applied for the first time to the derivation of the statistics of measurements with pre- and post-selection (\cite{ridley2023quantum}). In what follows we will construct quantum histories out of \emph{fixed points} in time which lie on both branches - recall that in this context, a `fixed point' occurs at a point in time where the state has equal forwards and backwards components. We therefore refer to it as the \emph{fixed point formulation} (FPF) of quantum mechanics. Because this picture allows the existence of many histories, it can essentially be regarded as a time-symmetric formulation of the Everett interpretation, or a consistent histories picture with time symmetric branching.

\subsection{The time contour}

\begin{figure}
\centering
  \includegraphics[width=.9\linewidth]
  {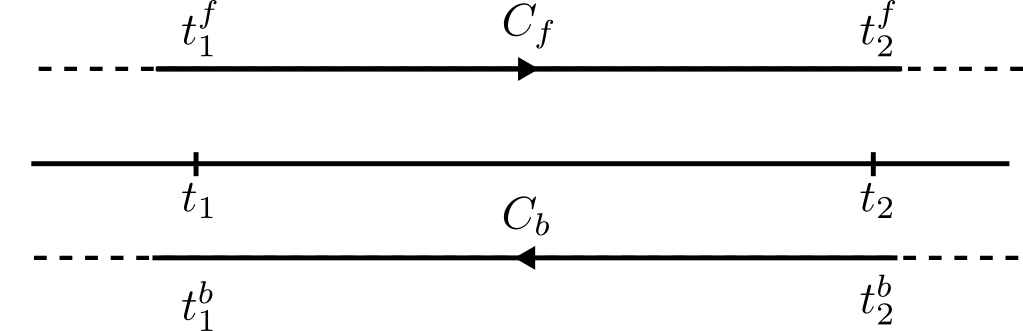}
  \caption{The Keldysh time contour in the time interval $\left[t_{1},t_{2}\right]$.}
  \label{fig:Keldysh_Contour}
\end{figure}

The Keldysh or nonequilibrium Green's function (NEGF) formalism is used to evaluate time-dependent expectation values of quantum observables $O\left(t_{2}\right)$ propagated from some initial time $t_{1}$:

\begin{equation}
 O\left(t_{2}\right) =  \mbox{Tr}\left[\rho_{1}U\left(t_{1},t_{2}\right)\hat{O}\left(t_{2}\right)U\left(t_{2},t_{1}\right)\right],
  \label{eqn:O_t}
\end{equation}  

where $\rho_{1}$ is the density matrix at $t_{1}$ and $U\left(t_{2},t_{1}\right)$ is the unitary evolution between times $t_{1}$ and $t_{2}$. The expression in Eq. \eqref{eqn:O_t} can be evaluated via two separate propagations, the first running forwards in time from $t_{1}$ to $t_{2}$, at which the operator $\hat{O}$ acts, before the system is propagated backwards from $t_{2}$ to $t_{1}$. This can be mapped to a propagation along the Keldysh time contour shown in Fig. \ref{fig:Keldysh_Contour}. The Keldysh contour consists of an `upper' branch $C_{f}$ of times $t^{f}$ on which the wavefunction is directed forwards in time, and a `lower' branch $C_{b}$ of times $t^{b}$ on which the direction of time is reversed. In particular, it is not possible to neglect the forwards or backwards part of the time propagation in the `future' or `past' of some measurement time $t$ without losing essential physical information.

In \cite{ridley2023quantum}, a similar physical state space to the ETNU space $\mathcal{H}\left(N_{t}\right)$ in Eq. \eqref{eq:ETNU} was proposed, with time-localized subspaces located at points along the Keldysh contour. We replicate the state space of the system at each time \emph{and} at each time orientation, much as in the consistent histories framework or the multiple-time state formalism. This is done by assigning a distinct Hilbert space to each time, $\mathcal{H}_{t_{i}}^{\alpha}$, where $\alpha\in\left\{ f,b\right\}$ denotes the upper or lower branch of $C$. Then, denoting contour position by the variable $z$, we can define a Fock-like space of events:

\textbf{Definition (Contour Space)} 

\begin{gather}\label{event_space}
    \mathcal{H}_{C}=\mathbb{C}\overset{\infty}{\underset{N_{t}=1}{\oplus}}\overset{N_{t}}{\underset{i=1}{\otimes}}\int_{C}^{\oplus}\mathcal{H}_{z_{i}}dz_{i}\\
    =\mathbb{C}\oplus \int_{C}^{\oplus}\mathcal{H}_{z_{1}}dz_{1}\oplus\int_{C}^{\oplus}\mathcal{H}_{z_{1}}\otimes\mathcal{H}_{z_{2}}dz_{1}dz_{2}\oplus\ldots
\end{gather}

In Eq. \eqref{event_space}, the symbol $\int_{C}^{\oplus}$ denotes a direct integral over all times on $C$ \cite{wils1970direct}. The universal wavefunction is defined on $\mathcal{H}_{C}$ in a representation-independent way, as a summation over 0-time, 1-time, 2-time...structures:

\textbf{Definition (Universal wavefunction)} 

\begin{gather}\label{universal_wave}
    \left|\Psi_{U}\right\rangle=\left|0\right\rangle+\overset{\infty}{\underset{N_{t}=1}{\Sigma}}\overset{N_{t}}{\underset{i=1}{\otimes}}\int_{C}\left|\psi_{i}\right\rangle dz_{i}
\end{gather}

Thus, the core ontology of the theory is provided by a multiple-time wavefunction defined on an \emph{unspecified} number of time points.

Now, given an ordering of $N_{t}$ times $t_{N_{t}}>t_{N_{t}-1}>...>t_{1}$, there are two corresponding ways of putting the times in order, one on each branch of $C \equiv C_{b} \oplus C_{f}$:

\begin{gather}\label{eqn:f_b_ordering}
t_{N_{t}}^{f}>_{C}t_{N_{t}-1}^{f}>_{C}\ldots>_{C}t_{1}^{f}\\
t_{N_{t}}^{b}<_{C}t_{N_{t}-1}^{b}<_{C}\ldots<_{C}t_{1}^{b}
\end{gather}

where the contour-ordering notation $>_{C}$, $<_{C}$ is introduced as in \cite{stefanucci_nonequilibrium_2013}. This is the main innovation of the Keldysh contour: ordering in time is distinct from the direction of dependence, since at a mathematical level the dependencies are oriented in the antichronological direction on the lower branch $C_{b}$. 

Each of the $N_{t}$ times in a history possesses two associated Hilbert spaces for the $f$ and $b$ components. Hence, the universal wavefunction has $2N_{t}$ temporal degrees of freedom and is a member of the \textit{contour} Hilbert space:
\begin{equation}
  \mathcal{H}_{C}=\mathcal{H}_{t_{N_{t}}}^{b}\otimes\mathcal{H}_{t_{N_{t}}}^{f}\otimes\ldots\otimes\mathcal{H}_{t_{1}}^{b}\otimes\mathcal{H}_{t_{1}}^{f}
  \label{eqn:Hilbert_Space}
\end{equation}  

A wavefunction in this space is not specified as a function of a single background time parameter. Rather, it is a multiple-time object built out of a sequence of moments. These moments consist of oppositely-oriented  temporal parts acting as `source' or `sink' states for processes on the branches $C_{f}$ and $C_{b}$. This codifies an eternalist view on time - all times in the future, past and present are incorporated equivalently into the theory within a block universe model of reality.

We make a corresponding first postulate:

\textbf{Ontological postulate}

\textit{The universal wavefunction $\left|\Psi_{U}\right\rangle \in \mathcal{H}_{C}$ for a history sequence of $N_{t}$ times is a `stack' of $2N_{t}$ temporal parts with fixed ordering on $C$, dividing time into $2(N_{t}-1)$ separate regions:}
\begin{equation}
\left|\Psi_{U}\right\rangle =\bigotimes_{i=1}^{N_{t}}\left|\Psi^{b}\left(t_{i}^{b}\right)\right\rangle \otimes\left|\Psi^{f}\left(t_{i}^{f}\right)\right\rangle 
\label{eqn:Psi_Universe}
\end{equation}

Here, $\left|\Psi^{\alpha}\left(t_{i}^{\alpha}\right)\right\rangle$  is restricted to the $C_{\alpha}$ time branch, and in general $\left|\Psi^{f}\left(t_{i}^{f}\right)\right\rangle\neq\left|\Psi^{b}\left(t_{i}^{b}\right)\right\rangle$. The inner product is defined on the Hilbert space $\mathcal{H}_{t_{i}}^{\alpha}$ in the usual way, such that $\left\langle \Psi_{U}\right|\left.\Psi_{U}\right\rangle =1$, which implies $\left\langle \Psi^{\alpha}\left(t_{i}^{\alpha}\right)\right|\left.\Psi^{\alpha}\left(t_{i}^{\alpha}\right)\right\rangle =1$ for any $\alpha$. Oppositely-oriented parts of the wavefunction are connected independently on $C_{f}$ and $C_{b}$. 

We now introduce the second core postulate, encoding the transfer of physical information between states defined on $C$:

\textbf{Dynamical postulate}

\textit{The time derivative of the wavefunction at each point on $C$ is given by the TDSE:}
\begin{equation}
i\hbar\partial_{t^{\alpha}}\left|\Psi^{\alpha}\left(t^{\alpha}\right)\right\rangle =H^{\alpha}\left(t^{\alpha}\right)\left|\Psi^{\alpha}\left(t^{\alpha}\right)\right\rangle 
\label{eqn:Branch_TDSE}
\end{equation}

Note that in every case of physical propagation (in the theory of full counting statistics, a non-physical auxiliary counting field is introduced, breaking \textbf{time symmetry} - see \cite{tang_full-counting_2014,ridley_numerically_2018}) the Hamiltonian operator is branch-independent, i.e. it takes on values on the upper/lower branches which are equal for the same physical time, $H^{b}\left(t^{b}\right)=H^{f}\left(t^{f}\right)$, and thus the dynamics respects \textbf{time symmetry}. For simplicity, indices on time arguments can therefore be dropped, $\left|\Psi^{\alpha}\left(t^{\alpha}\right)\right\rangle\equiv\left|\Psi^{\alpha}\left(t\right)\right\rangle$. 

The TDSE in Eq. \eqref{eqn:Branch_TDSE} defines a unitary mapping $U^{\alpha}\left(t_{2},t_{1}\right):\mathcal{H}_{t_{1}}^{\alpha}\mapsto\mathcal{H}_{t_{2}}^{\alpha}$ between the Hilbert spaces of different times on a single branch $\left|\Psi^{\alpha}\left(t_{2}\right)\right\rangle =U^{\alpha}\left(t_{2},t_{1}\right)\left|\Psi^{\alpha}\left(t_{1}\right)\right\rangle$, where $U^{\alpha}\left(t_{2},t_{1}\right)\equiv U^{\alpha}\left(t_{2}^{\alpha},t_{1}^{\alpha}\right)$ has the form \cite{stefanucci_nonequilibrium_2013}
\begin{equation}
U^{\alpha}\left(t_{2},t_{1}\right)=\hat{T}_{C}\exp\left[-\frac{i}{\hbar}\int_{t_{1}^{\alpha}}^{t_{2}^{\alpha}}d\tau H^{\alpha}\left(\tau\right)\right]
\label{eqn:Branch_Propagator}
\end{equation}

and $\hat{T}_{C}$ orders operators chronologically (latest to the left) on $C_{f}$, and anti-chronologically on $C_{b}$.

\subsection{The Fixed-Point Formulation}

We are now in a position to formalize the notion of a \emph{fixed point} in contour space $\mathcal{H}_{C}$, which represents an event.

\textbf{Definition (Fixed Point) }

\textit{A fixed point at time $t$ is a temporal part of the wavefunction in the $\mathcal{H}_{t}^{b}\otimes\mathcal{H}_{t}^{f}$ subspace, with equal $f$ and $b$ parts.}

Given a preparation of the state $\left|\psi\right\rangle$ of a system at some time $t_{1}$, all quantum histories in $\left|\Psi_{U}\right\rangle$ consistent with this preparation are constrained regardless of the contour branch. So there is a fixed point state at $t_{1}$, which is denoted:
\begin{equation}
\left\llbracket \psi\right\rrbracket_{t_{1}} \equiv\left|\psi^{b}\left(t_{1}\right)\right\rangle \otimes\left|\psi^{f}\left(t_{1}\right)\right\rangle
\label{eqn:1FP_Compact}
\end{equation}

\begin{figure}
\centering
  \includegraphics[width=.85\linewidth]{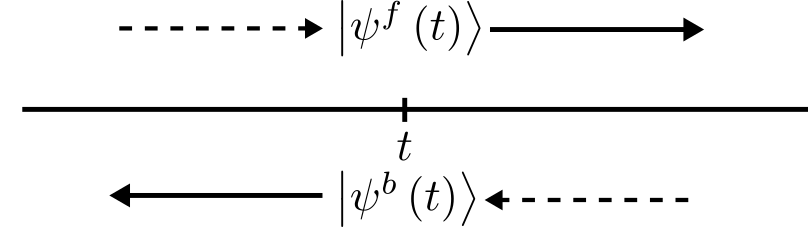}
  \caption{A single fixed point on the Keldysh contour.}
  \label{fig:1FP}
\end{figure}

This corresponds to an event in which the state is specified at $t_{1}$ (or a time-indexed projection, in the consistent histories language). In the terminology of NEGF, it corresponds to a `turning point' on the Keldysh contour at time $t_{1}$, i.e. to a point at which the time propagation along $C$ switches from the upper to the lower branch (\cite{stefanucci_nonequilibrium_2013}). We may think of the `present' time $t$ as `pinched' in between the upper-branch and lower-branch times $t^{f}$, $t^{b}$. The fixed point state connects to other points on $C$ in both time directions, in accordance with Eq. \eqref{eqn:Branch_TDSE}. 

We illustrate a fixed point on $C$ in Fig. \ref{fig:1FP}: the forward-directed part of the fixed point defined at $t$ is oriented towards times occurring `later' than $t^{f}$ on $C_{f}$, and the backward-directed part is oriented towards times occurring `later' than $t^{b}$ on $C_{b}$. Each fixed point is connected to four temporal regions: it acts as a `source' of wavefunction in both time directions (the thick black arrows on Fig. \ref{fig:1FP}), and a `sink' for parts of the wavefunction propagating from times lying `earlier' on $C$ (dashed lines on Fig. \ref{fig:1FP}). Thus, for a full description of a measurement connecting times across the region $\left[t_{1},t_{2}\right]$ at least two fixed points are required, $N_{t}\geq2$ in Eq. \eqref{eqn:Psi_Universe}. A quantum history sequence is defined in these terms:

\textbf{Definition} (Quantum history)

\textit{A quantum history $\left|h_{\mathbf{k}}\right\rangle$ extending across the time range $\left[t_{1},t_{2}\right]$ is a product state constructed from a sequence ${\mathbf{k}}=\left\langle k_{1},...,k_{N_{t}}\right\rangle $ of $N_{t}\geq2$ fixed points
\begin{equation}\label{eq:Quantum_History}
    \left|h_{\mathbf{k}}\right\rangle=\overset{N_{t}}{\underset{i=1}{\otimes}}\left\llbracket \psi_{k_{i}}\right\rrbracket _{t_{i}}
\end{equation}
connected by unitary mappings and bounded by fixed points at $t_{1}$ and $t_{2}$.}

In Eq. \eqref{eq:Quantum_History}, each $k_{i}$ in a history $\left|h_{\mathbf{k}}\right\rangle$ ranges over a complete basis set spanning $\mathcal{H}_{t_{i}}^{\alpha}$. To allow us to apply the usual rules of probabilistic reasoning to quantum histories, we define a \emph{family} of quantum histories $\mathcal{F}_{H}$ by imposing the consistency condition that any pair of histories in a family $\left\{ \left|h_{\mathbf{k}}\right\rangle \right\} $ must be non-overlapping:
\begin{equation}\label{eq:Quantum_History_Consistency}
    \left\langle h_{\mathbf{l}}\right.\left|h_{\mathbf{k}}\right\rangle =\delta_{\mathbf{kl}},
\end{equation}

where $\mathbf{k}\neq\mathbf{l}$ if $\left\llbracket \psi_{k_{i}}\right\rrbracket _{t_{i}}\neq\left\llbracket \psi_{l_{i}}\right\rrbracket _{t_{i}}$ for at least one value of $i \in \left[1,...,N_{t}\right]$. The consistency condition Eq. \eqref{eq:Quantum_History_Consistency} prevents the overlap of histories composed of different numbers of times $N_{t}$. 

Each set of quantum histories provides a set of distinct but complementary descriptions of the system over time, which may or may not correspond to measurement sequences. The fixed points constitute the time-localized `becomings' that the larger history sequences are composed of. They correspond to points in spacetime at which distinct quantum histories coincide, or \emph{crossing points} for quantum histories, where the future and past are constrained symmetrically. 

Now, following the terminology of \cite{vaidman_schizophrenic_1998}, the \emph{measure of existence} of a history may be defined as the relative size of the wavefunction region occupied by that history.

\textbf{Definition} (Measure of existence)

\textit{The measure of existence $m\left(h_{\textbf{k}}\right)$ of a quantum history $\left|h_{\mathbf{k}}\right\rangle$ containing $N_{t}$ fixed points in the time range $\left[t_{1},t_{2}\right]$, is the ratio of the integral of the wavefunction $\triangle\Psi_{\textbf{k}}$ along this history, to that of all histories 
\begin{equation}\label{MOE}
    m\left(h_{\textbf{k}}\right)=\frac{\ensuremath{\triangle\Psi_{\textbf{k}}}}{\underset{\mathbf{k'}}{\sum}\ensuremath{\triangle\Psi_{\textbf{k'}}}}
\end{equation}
in a family $\mathcal{F}_{H}$ consistent with the fixed point boundary conditions at $t_{1}$ and $t_{2}$.
}

Following on from this, we make the final postulate of the FPF:

\textbf{Statistical postulate (Vaidman rule)}
\textit{The quantum probability of a quantum history is equal to its measure of existence in the universal wavefunction.} 

The Vaidman rule is a conceptual postulate which allows the Born measure to be derived \cite{ridley2023quantum}. In doing so, it utilises a notion of probability which is based upon relative proportions of reality, which is assumed to be in one-to-one-correspondence with the wavefunction defined in eq. \ref{eqn:Psi_Universe}. The idea of the measure of existence was originally postulated for the many worlds interpretation of quantum mechanics \cite{vaidman_schizophrenic_1998,groisman_measure_2013}, and in that context there remains an ongoing debate over whether it provides an adequate solution to the problem of defining quantum probabilities (\cite{AdlamEverett,Thebault2015-DAWMWD}). It's possible that similar concerns would apply within the approach we suggest here as well, althugh we emphasize that the measure defined in Eq. \eqref{MOE} is directly grounded in physical ontology. We will not attempt to resolve these issues here  - for the purposes of this paper we will set conceptual difficulties aside and simply postulate the Vaidman rule. 

In \cite{ridley2023quantum}, it was proven that this postulate implies the correct \emph{mathematical} formalism in the case of a quantum measurement. Specifically, let $h_{\left\langle \psi,\phi\right\rangle }$ denote the quantum history corresponding to a measurement of $\left|\phi\left(t_{2}\right)\right\rangle$ at time $t_{2}$ following a preparation of the state $\left|\psi\left(t_{1}\right)\right\rangle$ at the initial time $t_{1}$. This process has the following measure of existence:
\begin{align}\label{eqn:Born_Rule}
m\left(h_{\left\langle \psi,\phi\right\rangle }\right)=\left|\left\langle \psi\left(t_{1}\right)\right|U\left(t_{1},t_{2}\right)\left|\phi\left(t_{2}\right)\right\rangle \right|^{2}
\end{align}

By the Vaidman rule, $m\left(h_{\left\langle \psi,\phi\right\rangle }\right)$ is just the quantum probability for this measurement process. Therefore this rule enables us to derive the Born probability measure. A similar derivation may be made for the ABL rule in Eq. \eqref{ABL} by considering quantum histories composed of three fixed points (\cite{ridley2023quantum}). These derivations are not based upon `rationality' or symmetry principles, but rather seek the mathematical form of the quantum probability as a structural feature of the underlying ontology of the theory, which is nothing less than the universal wavefunction. This is done via the identification of quantum probabilities with the relative sizes of wavefunction regions, in a construction which respects both \textbf{time symmetry} and \textbf{event symmetry}.

\subsection{`All-at-once' resolution of Maudlin's challenge}\label{allatonceMaudlin}

In this section we show how Maudlin's challenge can be circumvented within the FPF. However, we contend that the key to resolving Maudlin's challenge and related problems for retrocausal quantum mechanics \cite{maudlin2011quantum,bracken2021quantum} is to view fixed points as constraints on the future and past, not as dynamical pairs of `offer' and `confirmation' waves propagating through spacetime. Consecutive fixed points in a history influence each other via mutual causation - there is no sense in which one fixed point `causes' the other as the two directions of time are both needed along all points of the Keldysh contour. Thus, the FPF provides a representational model of reality in which an atemporal `all-at-once' retrocausality is instantiated, of the type described in \cite{adlam2022roads}.  We note that several proposed solutions to Maudlin's challenge exist in the literature(\cite{berkovitz2002causal,kastner2006cramer,marchildon2006causal}),  and in fact, as discussed by \cite{adlam2022roads}, they largely appear to be making use of something like `all-at-once' retrocausality, so the approach suggested here can be regarded as an alternative way of formulating this over-arching idea. 

Armed with the FPF, we now consider Maudlin's setup in Fig. \ref{fig:Maudlin_1} anew. Introducing the fixed point notation 

\begin{equation}\label{eq:LR}
    \left\llbracket L+R\right\rrbracket _{t_{0}}
\end{equation}

for the symmetrically separated photon state 

\begin{equation}\label{eq:LR_explicit}
    \left|L+R\right\rangle =\frac{1}{\sqrt{2}}\left(\left|L\right\rangle +\left|R\right\rangle \right)
\end{equation}

at time $t_{0}$, and 

\begin{equation}\label{eq:Detector}
    \left\llbracket P(A,B) \right\rrbracket _{t_{i}}
\end{equation}

for the fixed point state corresponding to the proposition $P(A,B)$ describing the detection of photons at detectors $A$ and $B$ by time $t_{i}$. Here, $P(A,B)=A \land B$ if and only if both detectors flash, $P(A,B)=A \land \neg B$ if and only if there is a detection event at $A$ but not at $B$, etc.. The physically possible quantum histories are then listed as follows:

\begin{align}
    \label{eq:Maudlin_Histories}
    \left|h_{\mathbf{1}}\right\rangle=\left\llbracket L+R\right\rrbracket _{t_{0}}\otimes\left\llbracket A \land \neg B \right\rrbracket _{t_{1}}\otimes\left\llbracket A \land  B \right\rrbracket _{t_{2}}\\
    \left|h_{\mathbf{2}}\right\rangle=\left\llbracket L+R\right\rrbracket _{t_{0}}\otimes\left\llbracket A \land \neg B \right\rrbracket _{t_{1}}\otimes\left\llbracket \neg A \land B \right\rrbracket _{t_{2}}\\
    \left|h_{\mathbf{3}}\right\rangle=\left\llbracket L+R\right\rrbracket _{t_{0}}\otimes\left\llbracket \neg A \land \neg B \right\rrbracket _{t_{1}}\otimes\left\llbracket A \land B \right\rrbracket _{t_{2}}\\
    \left|h_{\mathbf{4}}\right\rangle=\left\llbracket L+R\right\rrbracket _{t_{0}}\otimes\left\llbracket \neg A \land \neg B \right\rrbracket _{t_{1}}\otimes\left\llbracket \neg A \land B \right\rrbracket _{t_{2}}
\end{align}

Here, we physically rule out the fixed points corresponding to $P(A,B)=\neg A \land B$ at time $t_{1}$ because detector $A$ is placed in front of $B$. $P(A,B)= A \land B$ at time $t_{1}$ is ruled out by construction - photons cannot reach $B$ by this time. We rule out $P(A,B)= A \land \neg B$ at time $t_{2}$ since detection of a photon by $A$ means that $B$ is still located on the right hand side at $t_{2}$. We also rule out $P(A,B)= \neg A \land \neg B$ at time $t_{2}$ because if $A$ has not detected a particle by time $t_{2}$, $B$ must have swung around and detected it on the left-hand side of the apparatus. 

Next, we evaluate the measure of existence corresponding to each of the permitted histories $\left|h_{\mathbf{i}}\right\rangle$. We have 

\begin{equation}
    m\left(h_{\textbf{1}}\right)=\frac{1}{2}
\end{equation}

\begin{equation}
    m\left(h_{\textbf{2}}\right)=0
\end{equation}

\begin{equation}
    m\left(h_{\textbf{3}}\right)=0
\end{equation}

\begin{equation}
    m\left(h_{\textbf{4}}\right)=\frac{1}{2}
\end{equation}

Thus, we have two physically permitted quantum histories with non-zero measure, one corresponding to a right-travelling photon ($\left|h_{\mathbf{1}}\right\rangle$), the other to a particle travelling left ($\left|h_{\mathbf{4}}\right\rangle$). This resolves Maudlin's challenge as it is in line with the experimentally expected measurement statistics. In the FPF, both histories $\left|h_{\mathbf{1}}\right\rangle$ and $\left|h_{\mathbf{4}}\right\rangle$ are realised, on parallel trajectories through the universal wavefunction. These trajectories may be thought of as distinct time-extended Everettian worlds. 

\section{Conclusions}

In this article, we have argued that research on temporal symmetry in physics has largely focused on what we call \textbf{time symmetry}, neglecting another kind of symmetry principle that we call \textbf{event symmetry}. This observation is particularly relevant to the idea that retrocausality might have a role to play in making quantum mechanics more temporally symmetric, for it turns out that, perhaps surprisingly, models which simply incorporate ordinary forwards-in-time evolution can easily be expressed in a form that obeys \textbf{event symmetry} whereas this is quite non-trivial to achieve for approaches like the TSVF and the transactional interpretation that explicitly allow retrocausality. 

However, in this article we have shown that there is in fact a way to implement \textbf{event symmetry} in the context of an explicitly retrocausal model - this can be done by formulating multiple-time quantum states on the Keldysh contour, which improves upon the TSVF in a number of ways. The resulting model - the FPF - encodes a unitary and deterministic dynamics with no stochastic elements, and therefore may be interpreted as a time-symmetric version of Everettian quantum mechanics. The possibility remains, however, that this mathematical formalism could be interpreted in other ways, and we hope to investigate this in future work. It would also be interesting to consider further generalisations of the formalism - for example, the current approach assumes a global time-ordering or perhaps preferred foliation of spacetime, and it would seem natural to consider if this assumption could be relaxed to give an explicitly covariant model. 

Finally, we note that efforts to combine quantum mechanics and general relativity have thus far made gravity quantum mechanical - they have accepted indeterminism, a state-based ontology and a dynamical time evolution. We, by contrast, are making quantum mechanics more friendly to the Einsteinian picture - we implement determinism, an event based ontology (or at least an ontology that formalises the notion of an event), and an all-at-once block universe picture. This may have implications for unification (\cite{maccone2019fundamental,stoica2021post,iyer2022signatures,giovannetti2023geometric}).

\section*{Funding}

This work has been supported in part by the Israel Science Foundation Grant No. 2064/19 and the National Science Foundation–US-Israel Binational Science Foundation Grant No. 735/18.

\section*{Acknowledgements}

We are grateful to Huw Price, Riku Tuovinen and Ken Wharton for helpful remarks on a preprint of this paper.


\end{document}